\documentclass[a4paper]{aa}

\usepackage{graphicx,natbib}
\usepackage{txfonts}
\usepackage{color}
\usepackage[usenames,dvipsnames]{xcolor}

\newcommand{\bz}{$\langle B_\mathrm{z} \rangle$}
\newcommand{\bzm}{$|\langle B_\mathrm{z} \rangle|$}
\newcommand{\teff}{$T_{\rm eff}$}
\newcommand{\vs}{$v_{\rm e}\sin i$}
\newcommand{\rs}{$r_{\rm sp}$}
\newcommand{\kms}{km\,s$^{-1}$}

\newcommand{\figps}[1]{\resizebox{\hsize}{!}{\rotatebox{0}{\includegraphics{#1}}}}

\newcommand{\fifps}[2]{\centering\resizebox{#1}{!}{\includegraphics{#2}}}

\newcommand{\beq}{\begin{equation}}
\newcommand{\eeq}{\end{equation}}

\begin{document}
\title{Detectability of small-scale magnetic fields in early-type stars}

\author{O.~Kochukhov\inst{1}
   \and N.~Sudnik\inst{2}}

\institute{
Department of Physics and Astronomy, Uppsala University, Box 516, 75120 Uppsala, Sweden
\and
Sobolev Astronomical Institute, Saint Petersburg State University, Universitetskij pr. 28, Staryj Peterhof, Saint Petersburg 198504, Russia
}

\date{Received 27 March 2013 / Accepted 03 May 2013}

\titlerunning{Small-scale magnetic fields in early-type stars}
\authorrunning{O.~Kochukhov and N.~Sudnik}

\abstract
{
Strong, globally-organized magnetic fields are found for a small fraction of O, B, and A stars. At the same time, many theoretical and indirect observational studies suggested ubiquitous presence of weak localized magnetic fields at the surfaces of massive stars. However, no direct detections of such fields have been reported yet.
}
{
We have carried out the first comprehensive theoretical investigation of the spectropolarimetric observational signatures of the structured magnetic fields. These calculations are applied to interpret null results of the recent magnetic surveys of massive stars.
}
{
The intensity and circular polarization spectra of early-type stars are simulated using detailed polarized radiative transfer calculations with LTE model atmospheres. Similar to observational analyses, the mean Stokes $I$ and $V$ line profiles are obtained by applying a multi-line averaging technique. Different spectropolarimetric observables are examined for multiple realizations of randomly distributed radial magnetic field spots with different spatial scales.
}
{
We characterize the amplitude of the circular polarization profiles and the mean longitudinal magnetic field as a function of magnetic spot sizes. The dependence of these observables on the effective temperature, projected rotational velocity, and inclination angle is also investigated. Using results of the recently completed Magnetism in Massive Stars (MiMeS) survey, we derive upper limits on the small-scale magnetic fields compatible with the MiMeS non-detections.
}
{
According to our simulations, existing spectropolarimetric observations of sharp-lined massive stars rule out the presence of the small-scale fields stronger than 50--250~G, depending on the typical spot sizes. For broad-lined stars, the observations constrain such fields to be below approximately 1~kG.
}


\maketitle

\section{Introduction}
\label{intro}

There is a growing observational and theoretical support for a magnetic dichotomy among intermediate-mass and massive stars. A small fraction of the late-B and A stars possesses strong organized magnetic fields at their surfaces \citep[see reviews by][]{donati:2009,mathys:2009,kochukhov:2011c}. These magnetic field topologies are approximately dipolar for the majority of stars, often inclined relative to the stellar rotational axis, and have strengths of up to $\sim$\,30~kG. The global magnetic fields of intermediate-mass stars are believed to be stable fossil remnants of the magnetic flux acquired at the stellar formation phase \citep{braithwaite:2006}. These fields exhibit a lower threshold of $\sim$\,300~G, below which the large-scale magnetic topologies are probably not able to withstand the shearing by the differential rotation \citep{auriere:2007}.

The incidence of fossil stellar magnetic fields is now understood to extend well into the mass range of early-B and O stars. The Magnetism in Massive Stars \citep[MiMeS;][]{wade:2011} survey and other studies \citep{wade:2006b,hubrig:2011} have identified several examples of O stars with globally-organized, kG-strength magnetic fields. These discoveries are particularly frequent among the so-called Of?p stars \citep{martins:2010,wade:2011a,wade:2012}, which appear to represent a massive-star extension of the Ap/Bp phenomenon. Remarkably, magnetic fields existing on the surface of some of these peculiar O stars exceed 20~kG \citep{wade:2012a}, rivaling the strongest fields found in much smaller Ap/Bp stars.

In addition to the strong, large-scale magnetic fields found for a handful of massive stars there exists a substantial indirect observational evidence and credible theoretical predictions for the existence of localized, and probably weak, magnetic fields in the majority of massive stars. Much of the indirect support for such fields is coming from various poorly understood wind and surface variability phenomena. For example, nearly all O-type stars and many B stars show unexplained cyclical variability in their winds \citep{kaper:1996,fullerton:1996,prinja:2002}. The most prominent feature of this variability is discrete absorption components (DACs), which are observed to accelerate through the UV-wind line profiles and maintain coherency over several stellar rotations. These features are probably caused by corotating interacting regions (CIRs) in the line-driven stellar wind, as proposed by \citet{mullan:1984}. In this model, a faster moving stream collides with a slow moving one, leading to the formation of curved large scale structure in the stellar wind. The modeling by \citet{cranmer:1996} demonstrated that such wind structures give rise to DACs in the wind-sensitive UV line profiles. The velocity contrast required by this model arises from perturbations at the photospheric level, which may be connected to evolving small-scale magnetic spots.

An example of the compelling observational evidence of the connection between DACs and a magnetic field was provided by the long-term monitoring of the magnetic early-O star $\theta^1\,$Ori~C. \citet{henrichs:2005} showed that 13 years of the \ion{C}{iv} UV profile observations of this star, phase-folded with the 15.4~d rotational period, show a progressing absorption feature very similar to individual DACs observed in other early-type stars. $\theta^1\,$Ori~C is an oblique rotator with a stable  dipolar magnetic field \citep{donati:2002,wade:2006b}, so the feature in the UV wind lines reoccurs exactly at the same rotational phase, associated with one of the magnetic poles. Using this well-studied O-star as an example, one can suspect magnetic field to be the main culprit of the surface perturbations leading to DACs in other massive stars.

A considerable theoretical effort was devoted to investigating various alternatives to the fossil magnetism of early-type stars. Massive stars generate strong magnetic fields in their convective cores \citep{brun:2005}, but these fields do not reach the stellar surface on the observable time scales \citep{macgregor:2003}. A dynamo process operating in the differentially-rotating radiative stellar interiors \citep{spruit:2002} can also produce magnetic fields emerging on the stellar surface \citep{mullan:2005}. The reality of this particular magnetic field generation mechanism, which has important implications for the evolution of massive stars \citep{maeder:2005} and their descendants \citep{heger:2005}, was however disputed by \citet{zahn:2007}. Finally, recent stellar structure calculations of hot massive stars suggested the presence of sub-surface convectively unstable regions, related to the iron opacity peak \citep{cantiello:2009}. By analogy with the turbulent magneto-convection operating in other astrophysical environments \citep[e.g.][]{kapyla:2008}, these sub-surface convection zones (SCZ) can produce magnetic fields in the presence of shear and stellar rotation and provide an attractive framework for the explanation of other massive-star atmospheric phenomena, including excessive micro- and macroturbulent line broadening \citep{hunter:2008} and stochastic non-radial oscillations \citep{degroote:2010}. Preliminary MHD simulations by \citet{cantiello:2011a} confirmed the ability of SCZ to generate localized magnetic fields visible at the stellar surface. The properties of such fields and associated brightness spots were further investigated analytically by \citet{cantiello:2011}.

Simultaneously with these developments for massive stars, the discovery of sub-G magnetic fields in the brightest intermediate-mass stars Vega \citep{lignieres:2009} and Sirius \citep{petit:2011} hinted at the hidden magnetism of normal A and late-B stars. In response to these findings, \citet{braithwaite:2013} developed a theoretical model of the dynamically-evolving ``failed'' fossil magnetic fields, which are expected to be much weaker and structured on smaller scales than the well-known stable fossil fields of the magnetic Ap/Bp stars.

Despite an emerging interest in probing the small-scale magnetism of early-type stars and exploring the consequences for the stellar physics, the current theoretical models have not reached the level of sophistication and detail necessary for making meaningful predictions about the strengths and topologies of the localized magnetic fields in individual stars. Therefore, currently there exists a gap between observational constraints, e.g. detections of the polarization signatures for Sirius and Vega and non-detections of the magnetic fields for most O- and B-type stars included in the MiMeS survey, and theoretical predictions. In particular, it is unknown what direct observational signatures of small-scale fields one can expect in the polarization spectra of the early-type stars and how the amplitude and variability of these features change with the characteristics of magnetic field and with the stellar parameters.

Even the very ability of modern spectropolarimetric methods to detect localized magnetic features on the stellar surfaces is a matter of debate. On the one hand, a contribution of the small-scale fields to the disk-integrated polarization spectra and to the resulting magnetic observables might be strongly reduced by the cancellation of different field polarities. On the other hand, high-resolution studies of polarization signatures inside spectral line profiles partly overcomes this problem thanks to the rotational Doppler resolution of the stellar surface \citep{semel:1989}. The interplay of the cancellation vs. Doppler resolution and its dependence on the stellar rotation and sizes of magnetic features has not been systematically investigated.

The goal of our study is to address several of these issues through a comprehensive and realistic numerical simulations of the circular polarization spectra corresponding to the small-scale magnetic fields of different degree of complexity. In the following, we describe the methodology of the magnetic spectrum synthesis and multi-line analysis of the theoretical Stokes $V$ spectra (Sect.~\ref{methods}), present a summary of the simulation results for a wide range of stellar and magnetic field parameters (Sect.~\ref{results}) and then use these data to constrain the strength of possible small-scale stellar magnetic fields by comparing our predictions with the outcome of recent magnetic field surveys of massive stars (Sect.~\ref{disc}).

\section{Methods}
\label{methods}

Two main spectropolarimetric observational techniques are currently being used in the context of large-scale magnetic field surveys of early-type stars. The low-resolution circular spectropolarimetry, using e.g. the FORS1/2 instrument at the ESO VLT, detects magnetic field by measuring the mean longitudinal magnetic field in the wings of the hydrogen Balmer lines or in the unresolved blends of metal lines \citep{bagnulo:2002a,wade:2007,hubrig:2011,bagnulo:2012}. This technique is only sensitive to the line of sight magnetic field component integrated over the visible stellar hemisphere and is poorly suited to provide constraints on intermittent, small-scale magnetic structures. Furthermore, flexures and calibration problems of the Cassegrain-mounted polarimeters, such as FORS1/2, limit their usefulness for the search of very weak magnetic field \citep[see discussion in][]{bagnulo:2012}.

The second common magnetic field detection technique relies on obtaining high-resolution circular polarization spectra covering a wide wavelength range and applying a line-averaging technique to combine information from all suitable metal absorption features. This approach is adopted by the MiMeS survey for the analysis of circular polarization spectra obtained with ESPaDOnS, NARVAL, and HARPSpol spectropolarimeters \citep{grunhut:2009,alecian:2011a,neiner:2012}. The mean Stokes $V$ line profiles obtained with the least-squares deconvolution (LSD) method \citep{donati:1997} are particularly useful for the magnetic field diagnostic since they behave similarly to real spectral lines \citep{kochukhov:2010a} and allow one to characterize and model spectrally-resolved Zeeman signatures. The LSD profiles are therefore sensitive to the field topologies yielding very low or null mean longitudinal fields \citep[e.g.][]{shultz:2012} and have been successfully used to detect and map complex dynamo-generated magnetic fields of cool active stars \citep{donati:2003,kochukhov:2013}. 

The goal of our study is to characterize observational signatures of the magnetic fields which are at least as complex topologically as the surface magnetic structures found for cool active stars. Hence, the high-resolution spectropolarimetry combined with a multi-line analysis appears to be the most suitable methodology for our investigation. More specifically, we simulate the LSD Stokes $I$ and $V$ profiles for magnetic field geometries of varying degree of complexity and study these polarization profiles as a function of the magnetic and stellar parameters. The following subsections describe computation of the local LSD line profiles, explain how we generate random magnetic field maps and perform surface integration and, finally, discuss derivation of different spectropolarimetric observables from the resulting disk-integrated polarization profiles.

\subsection{Local polarized line profiles}

The analysis of the LSD profiles of early-type stars presented in previous spectropolarimetric studies relied on a simplified analytical description of the intensity and circular polarization profile shapes. The most common assumption is to describe Stokes $I$ as a Gaussian and Stokes $V$ as its scaled derivative \citep[e.g.][]{petit:2012}. Although straightforward to implement, this approach suffers from ambiguities in the choice of an analytical line profile model and its parameters, and is unable to properly describe the temperature dependence of the local line profile characteristics. Moreover, averaging over a small number of diagnostic lines available in the optical spectra of early-type stars does not necessarily yield a simple, symmetric Stokes $I$ profile that could be described by an analytical function. Here we avoid these problems by numerically deriving the local LSD profiles from detailed polarized radiative transfer calculations of individual spectral lines.

The local Stokes $I$ and $V$ line profile tables were prepared by first computing the full local spectra covering the 4000--8000~\AA\ wavelength range with the {\sc synmast} polarized radiative transfer code and then applying the LSD line averaging procedure. A detailed description of the relevant computer codes and methodology required for this analysis can be found in our previous publications \citep{kochukhov:2010a}. 

The spectrum synthesis calculations were performed for five {\sc atlas9} LTE model atmospheres \citep{kurucz:1993c} with \teff\ in the range from 10\,000 to 30\,000~K with a step of 5000~K. For all models we adopted the surface gravity of $\log g$\,=\,4.0 and microturbulent velocity of 2~\kms. The atomic line data were extracted from the VALD database \citep{kupka:1999}, using the 1\% line depth selection threshold and the same microturbulent velocity. The spectral regions containing the hydrogen and helium lines were subsequently excluded from the input line lists. 

The local Stokes $IV$ parameters and the unpolarized continuum intensity $I_{\rm c}$ were computed for a grid of 20 limb angles $\theta$, distributed equidistantly with $\mu=\cos{\theta}$, and for the two values of local magnetic field strength, $B=0$ and 0.5~kG. For the radial magnetic field geometries considered below, the line of sight magnetic field component required for the polarized spectrum synthesis is given by $B_{\rm z}=B \mu$, independently of the distribution of magnetic features over the stellar surface. For all spectrum synthesis calculations we employed a microtubulence of 2~\kms, consistently with the choice of model atmospheres and the VALD line list extractions.

\begin{figure}[!th]
\centering
\figps{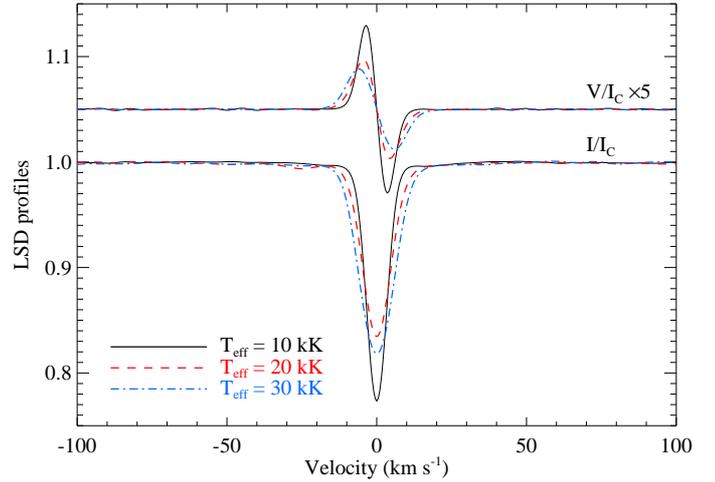}
\caption{Normalized disk-center LSD Stokes $I$ and $V$ profiles for different effective temperatures and a line-of-sight magnetic field of 0.5~kG. The Stokes $V$ spectra are offset vertically by 1.05 and expanded by a factor of 5.}
\label{fig:prf_teff}
\end{figure}

On the next step, the local Stokes $I(\lambda,\mu,B)$ and $V(\lambda,\mu,B)$ spectra were normalized by the corresponding continuum intensity $I_{\rm c}(\lambda,\mu)$ and convolved with the instrumental profile corresponding to the resolution of $R=65\,000$, appropriate for the ESPaDOnS and NARVAL spectropolarimeters. The resulting spectra were converted to the local LSD profiles, $I_{\rm LSD}(v,\mu,B)$ and $V_{\rm LSD}(v,\mu,B)$, by applying the least-squares deconvolution procedure to a subset of spectral lines deeper than 10\% of the continuum.  Simultaneously, the mean continuum intensity $\overline{I_{\rm c}}(\mu)$ was established by averaging $I_{\rm c}(\lambda,\mu)$ in the central wavelengths of the spectral lines included in the LSD masks. The LSD profiles were computed on a fine velocity grid with a step of 0.5~\kms\ covering the $\pm500$~\kms\ velocity range to ensure an accurate surface integration for \vs\ of up to 250~\kms.

Table~\ref{tab1} summarizes statistics of the line lists used for the spectrum synthesis and LSD profile calculations. Depending on the temperature, we used 1500--2600 lines for the spectrum synthesis and up to $\approx$\,400 lines for LSD. The mean wavelength $\lambda_0$, effective Land\'e factor $\overline{g}_0$, and line depth $d_0$ of the LSD masks are reported in the last three columns of Table~\ref{tab1}. The same mean parameters were used to renormalize the LSD weights for each \teff\ as described by \citet{kochukhov:2010a}.

Examples of the local disk-center LSD Stokes $I$ and $V$ profiles for \teff\,=\,10\,000, 20\,000, and 30\,000~K are presented in Fig.~\ref{fig:prf_teff}. The strongest LSD profiles are obtained for low \teff. The line width systematically increases towards higher temperatures. The local Stokes parameter line shapes are not exactly symmetric and contain weak features outside the main profiles due to blends unaccounted for in the LSD mask and due to limitations of the line averaging procedure itself. These LSD profile distortions are, of course, also present in observations but cannot be described by an analytical line profile model.

\begin{table}
\caption{Characteristics of the line lists used for synthetic LSD profile calculations.}
\label{tab1}
\centering
\begin{tabular}{c c c c c c c }
\hline\hline
\teff\ (K) & $N_{\rm tot}$ & $N_{\rm LSD}$ & $\lambda_0$ (\AA) & $\overline{g}_0$ & $d_0$ \\
\hline
10\,000  & 2609 & 428 & 4807 & 1.179 & 0.281 \\
15\,000  & 1542 & 196 & 5251 & 1.134 & 0.202 \\
20\,000  & 1473 & 208 & 4914 & 1.221 & 0.177 \\
25\,000  & 1698 & 331 & 4723 & 1.214 & 0.194 \\
30\,000  & 1501 & 258 & 4582 & 1.188 & 0.195 \\
\hline
\end{tabular}
\end{table}

\begin{figure*}[!t]
\centering
\fifps{8.5cm}{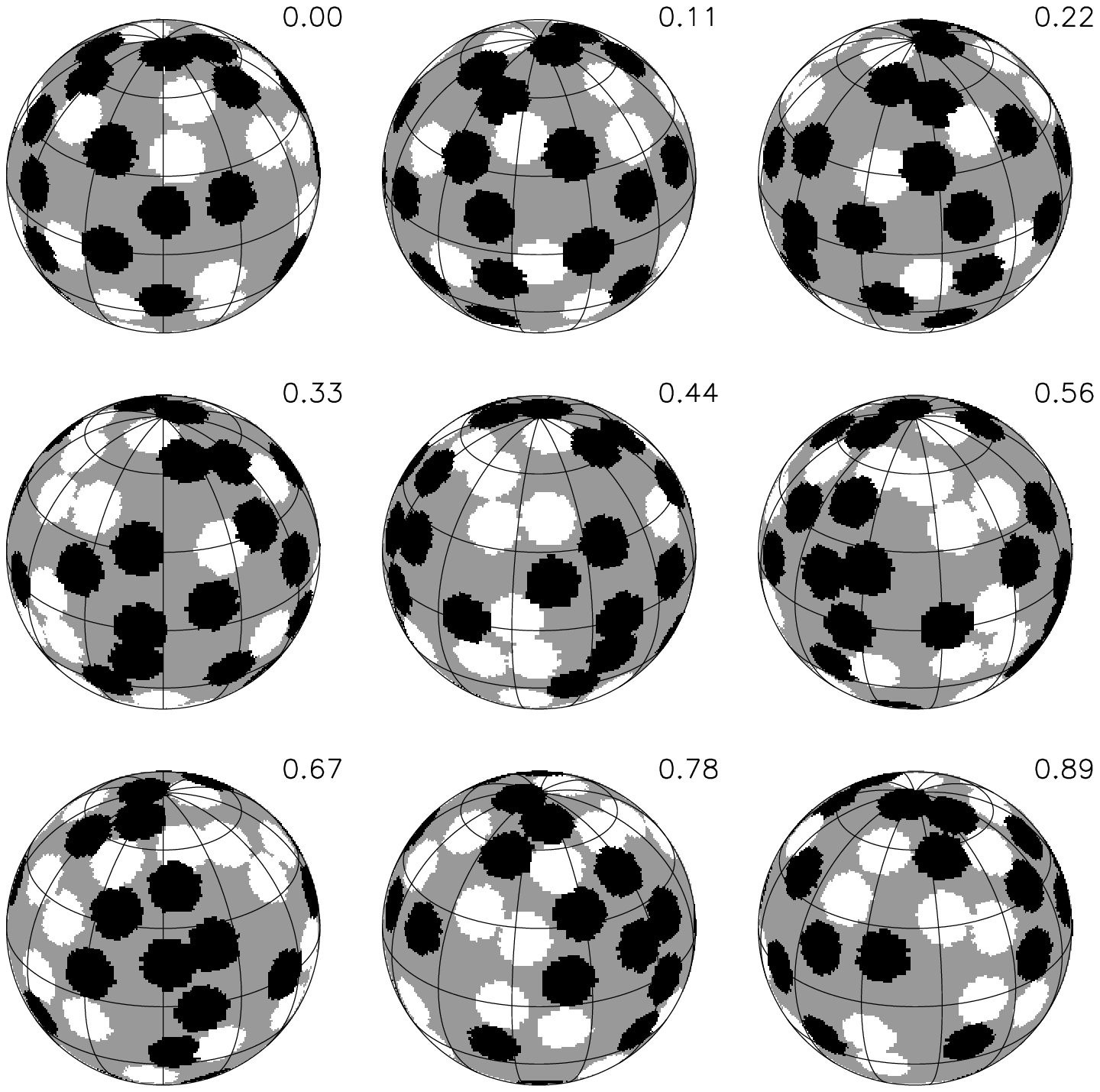}\hspace*{0.5cm}
\fifps{8.5cm}{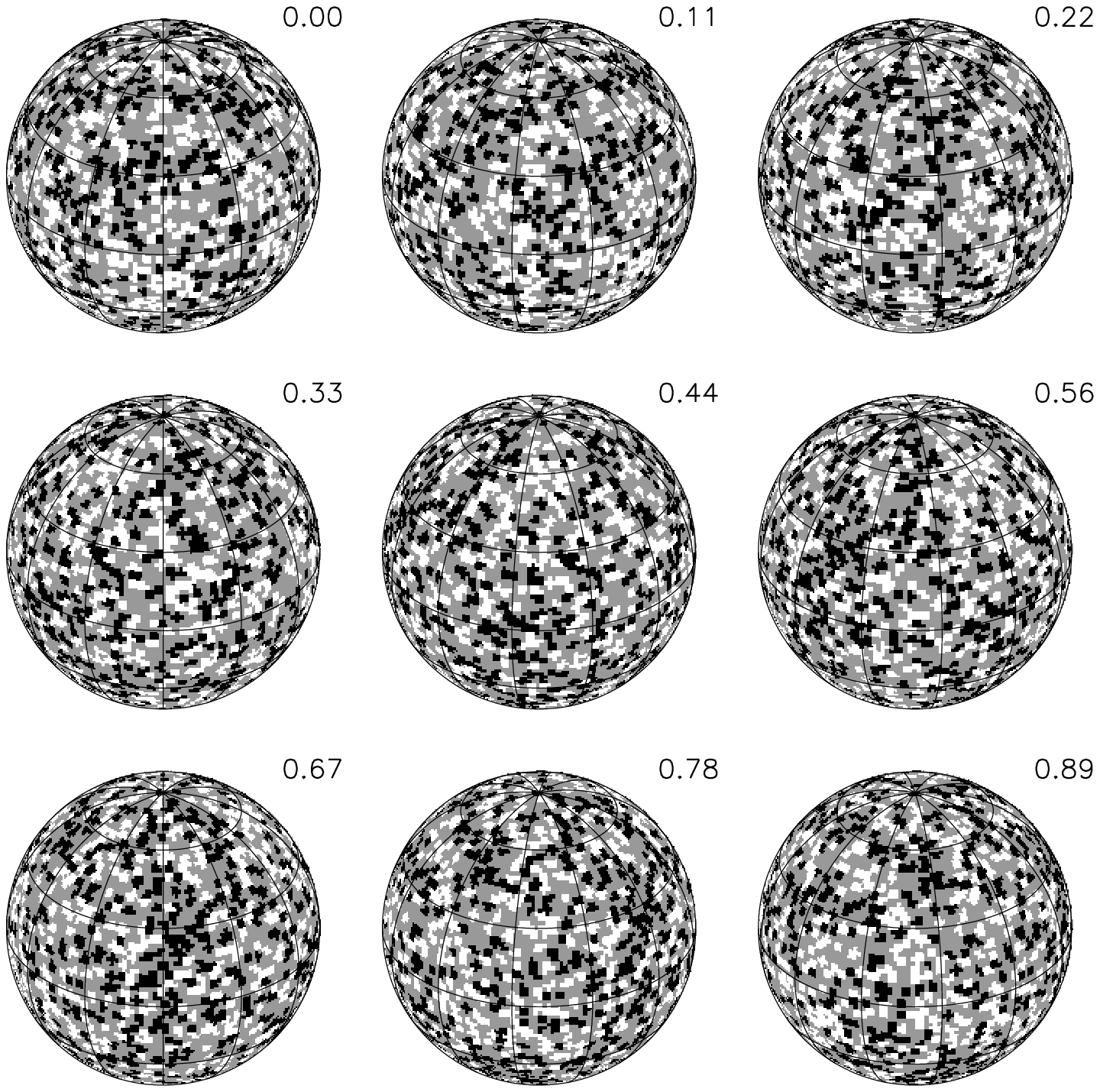}
\caption{Examples of random distributions of radial magnetic field spots with filling factor $f_{\rm sp}=0.5$ and \rs\,=\,10\degr\ (\textit{left panel}) and 2\degr\ (\textit{right panel}). In each case the star is shown at nine rotational phases and inclination angle $i=60\degr$. The white and black regions correspond to the spots with positive and negative field polarity, respectively. The grey areas are field-free.}
\label{fig:maps}
\end{figure*}

\subsection{Disk-integrated spectra for random magnetic fields}

Many different mathematical models can be used to construct small-scale magnetic field distributions on the stellar surfaces. Here we employ a simple formulation, requiring few arbitrary parameters but still maintaining sufficient physical realism. In the following we assume that the stellar magnetic field topology is given by a superposition of randomly distributed circular spots. The field inside the spots is purely radial, by analogy with sunspots, and is characterized by a single value of the magnetic field modulus. Outside the spots the photosphere is assumed to be non-magnetic. An equal number of spots with the outward- and inward-directed radial field is used, thus maintaining the zero net magnetic flux through the stellar surface. The average properties of such magnetic field distributions are defined by three parameters: the spot angular radius $r_{\rm sp}$, the field strength $B_{\rm sp}$ inside the spots, and the fraction of the stellar surface $f_{\rm sp}$ covered by the magnetic field (a filling factor).

\begin{figure}[!th]
\centering
\fifps{8cm}{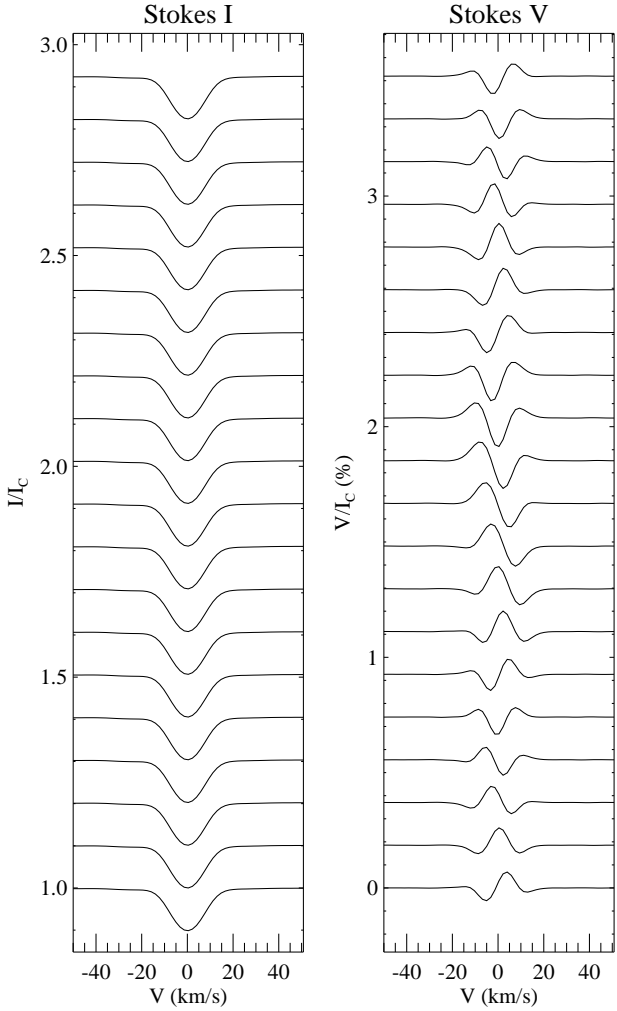}
\caption{Disk-integrated normalized LSD Stokes $I$ and $V$ profiles for \vs\,=\,10~\kms\ and a random distribution magnetic spots with \rs\,=\,10\degr. Other parameters are \teff\,=\,20\,000~K, $i=60\degr$, and magnetic filling factor $f_{\rm sp}=0.5$. The spectra corresponding to different rotational phases are offset vertically.}
\label{fig:prf_v10}
\end{figure}

\begin{figure}[!th]
\centering
\fifps{8.15cm}{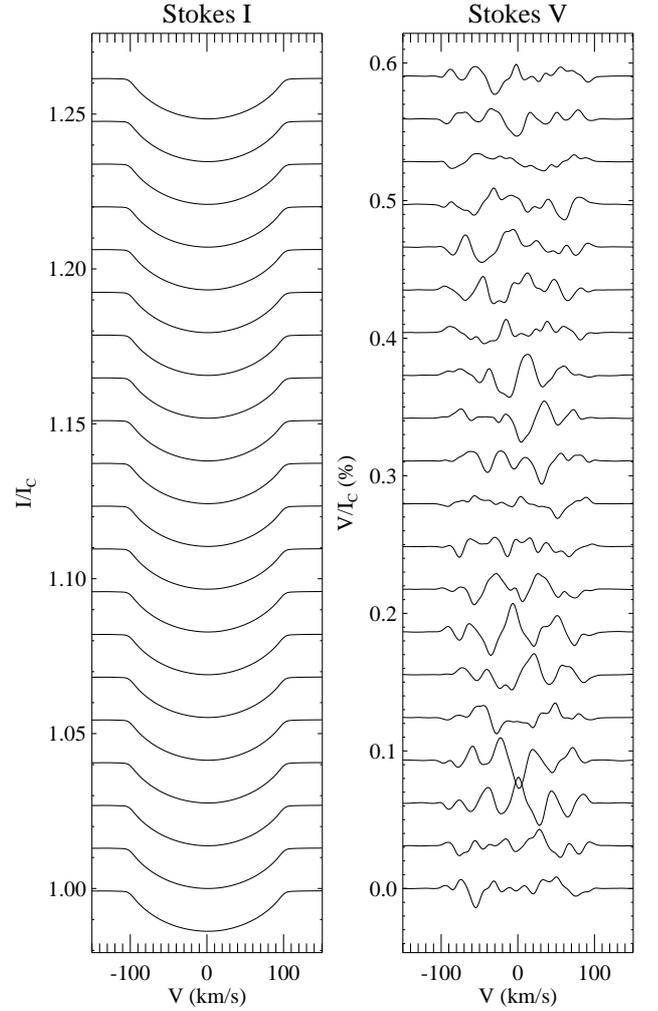}
\caption{LSD Stokes $I$ and $V$ profiles computed with the same magnetic map and setup as in Fig.~\ref{fig:prf_v10}, except for \vs\,=\,100~\kms.}
\label{fig:prf_v100}
\end{figure}

Once $r_{\rm sp}$ is set, the area of each spot can be computed as
\beq
a_{\rm sp} = 2\pi R^2_{\star} (1 - \cos{r_{\rm sp}})
\eeq
and the total number of spots for a given $f_{\rm sp}$ can be estimated by rounding
\beq
n_{\rm sp} = 4\pi R^2_{\star} f_{\rm sp} / a_{\rm sp}
\eeq
to the nearest even integer. Then, the latitudes $\chi \in [-\pi/2,\pi/2]$ and longitudes $\varphi \in [0,2\pi]$ of the spot centers are obtained from
\beq
\begin{array}{lcl}
\chi & = & \pi / 2 - \arccos{(2 x - 1)}\\
\varphi & = & 2\pi x^\prime,
\end{array}
\eeq
where $x$ and $x^\prime$ represent two different sequences of uniformly distributed random numbers in the range between 0 and 1. This algorithm of the random spot position generation may produce overlapping spots. In practice, we allowed the spots to overlap by no more than 10\% of their area, rejecting spot coordinates too close to the already existing spots.

The local Stokes $I$ line profiles in Fig.~\ref{fig:prf_teff} have a full-width at half maximum of $\delta v$\,=\,8--14~\kms. The corresponding approximate theoretical spatial resolution limit can be estimated as
\beq
r_{\rm min} = \frac{\displaystyle\pi}{4}\frac{\displaystyle\delta v}{v_{\rm e}\sin i}.
\eeq
Using $\delta v$\,=\,10~\kms, we can simplify this expression to $r_{\rm min}=900/v_{\rm e}\sin i$. This indicates that an isolated surface feature with $r_{\rm sp}=4.5\degr$ is fully resolved for \vs\,=\,200~\kms. Using this estimate as an guideline, we considered the spot sizes 2, 5, 10, 20, and 40\degr. An intermediate magnetic filling factor $f_{\rm sp}=0.5$ was adopted for the majority of calculations. The corresponding total spot numbers varied from 4 (for $r_{\rm sp}=40\degr$) to 1642 (for $r_{\rm sp}=2\degr$). 

Numerically, magnetic maps were evaluated on the spherical surface grid containing 15130 zones of roughly equal areas \citep[see Fig.~5 in][]{piskunov:2002a}. This fine surface grid was required to properly resolve the spots with the smallest spatial scales. An example of the random magnetic spot distributions computed for $r_{\rm sp}=10$ and 2\degr\ is illustrated in Fig.~\ref{fig:maps}.

After establishing a surface magnetic field map, we computed the disk-integrated Stokes $I$ and $V$ LSD profiles, $\langle I_{\rm LSD}\rangle$ and $\langle V_{\rm LSD}\rangle$, by appropriate interpolation and weighted summation of the Doppler-shifted local LSD Stokes profiles for a specific inclination angle $i$ and projected rotational velocity \vs. The line profiles were simulated for 20 equidistant rotational phases, using a velocity grid with 1.8~\kms\ step size. All calculations presented below were performed for the \vs\ values of 10, 20, 50, 100, and 200~\kms.

An example of the phase variation of the circular polarization profiles for the random magnetic field distribution with $r_{\rm sp}=10\degr$ and $f_{\rm sp}=0.5$ (the same as shown in the left panel of Fig.~\ref{fig:maps}) is given in Fig.~\ref{fig:prf_v10} and \ref{fig:prf_v100}. These figures illustrate the line profiles computed for identical surface maps and the same set of viewing geometries, but different projected rotational velocities. As expected, a higher projected rotational velocity (Fig.~\ref{fig:prf_v100}) reveals much more complex circular polarization profile shapes thanks to the Doppler resolution of the stellar surface. At the same time, the amplitude of the Stokes $V$ signatures is also considerably reduced compared to low the \vs\ case (Fig.~\ref{fig:prf_v10}).

All our calculations were performed for the 0.5~kG field strength inside magnetic spots. Since here we deal with the magnetic line formation in the weak-field regime, it can be assumed that the Stokes $V$ profiles and all resulting spectropolarimetric observables scale linearly with the magnetic field strength. Consequently, after verifying this behavior for the local Stokes $V$ LSD profiles, we did not vary the field strength inside magnetic spots in any of the numerical experiments presented in this paper.

\subsection{Magnetic observables}

The main goal of our study is to assess the capability of recent high-resolution spectropolarimetric surveys of early-type stars to detect intermittent magnetic fields on their surfaces. Consequently, we consider the amplitude of the Stokes $V$ signatures resulting from our numerical line profiles simulations as the primary magnetic field observable. For each set of circular polarization profiles we calculated the mean peak-to-peak Stokes $V$ amplitude $V_{\rm mean}$ by averaging the instantaneous amplitude over 20 rotational phases. We also determined the maximum amplitude $V_{\rm max}$ by finding the rotational phase with the largest circular polarization signature. To obtain statistically meaningful results with our random magnetic spot models, we averaged both $V_{\rm mean}$ and $V_{\rm max}$ over 100 different random realizations of the magnetic field distributions for the same $r_{\rm sp}$ and $f_{\rm sp}$. These calculations were repeated for a grid of five $r_{\rm sp}$ and five \vs\ values (see above), yielding 50\,000 individual synthetic line profiles for each \teff, $f_{\rm sp}$, $i$ combination.

In addition to considering the amplitude of the Stokes $V$ signatures, we introduced another parameter $n_{\rm z}$ characterizing complexity of the circular polarization profiles. This (integer) number was established by counting how many times a given Stokes $V$ profile intersects the zero line. This parameter also roughly corresponds to the number of lobes in the Stokes $V$ profile minus one. The final mean $n_{\rm z}$ reported below for each $r_{\rm sp}$-\vs\ combination was obtained by computing the median over 20 rotational phases and 100 random map realizations. We found that this simple diagnostic exhibits a smooth variation with stellar parameters and readily distinguishes complex and simple polarization signatures. For example, for the magnetic spot distribution with $r_{\rm sp}=10\degr$ illustrated in Fig.~\ref{fig:maps} $n_{\rm z}$ gradually increases from 2 for \vs\,=\,10~\kms\ (line profiles in Fig.~\ref{fig:prf_v10}) to 9 for \vs\,=\,100~\kms\ (line profiles in Fig.~\ref{fig:prf_v100}). 

\begin{figure*}[!t]
\centering
\figps{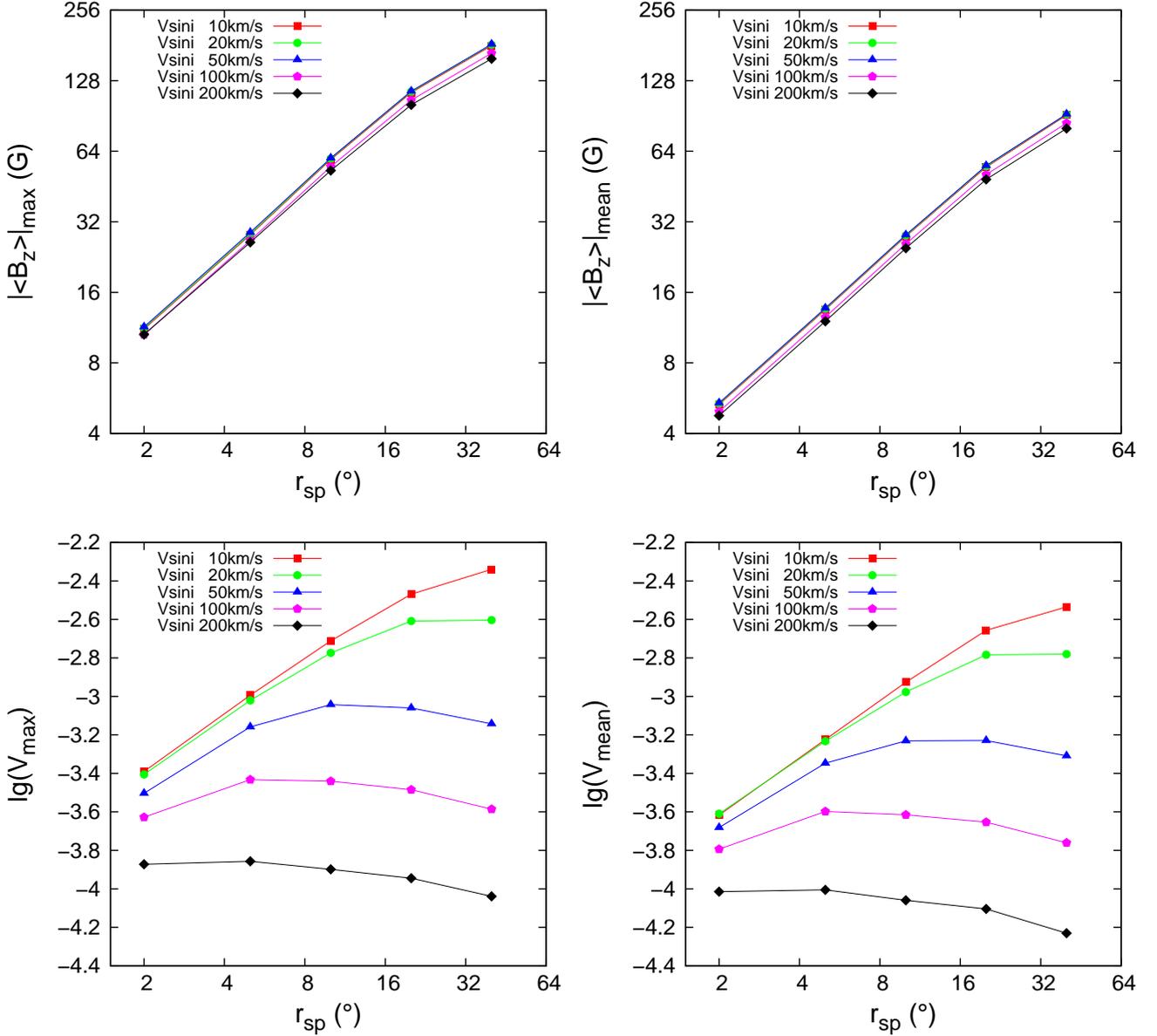}
\caption{Mean longitudinal magnetic field and the LSD Stokes $V$ profile amplitude as a function of the magnetic spot radius and projected rotational velocity. These calculations were carried out for \teff\,=\,20\,000~K and magnetic filling factor $f_{\rm sp}=0.5$. The top panels show maximum \bzm\ for a sample of 20 rotational phases (\textit{top left}) and \bzm\ averaged over this set of rotational phases (\textit{top right}). The bottom panels illustrate the corresponding maximum (\textit{bottom left}) and mean (\textit{bottom right}) circular polarization amplitude. Each point in this plot was computed by averaging over 100 different random realizations of the magnetic spot distributions.}
\label{fig:res1}
\end{figure*}

For completeness, we also determined the mean longitudinal magnetic field \bz\ from each pair of the disk-integrated Stokes $I$ and $V$ profiles. This was accomplished using the equation
\beq
\langle B_{\rm z} \rangle = - 7.145\times 10^6 \frac{\int v \langle V_{\rm LSD} \rangle \mathrm{d}v}{\lambda_0 \overline{g}_0 \int ( 1 - \langle I_{\rm LSD} \rangle) \mathrm{d}v},
\eeq
where $\lambda_0$ in \AA\ and the resulting \bz\ is in G. The mean wavelength $\lambda_0$ and effective Land\'e factor $\overline{g}_0$ are the same parameters as applied above for normalizing the LSD line weights (see Table~\ref{tab1}). Similar to the analysis of the Stokes $V$ amplitude, the maximum and mean unsigned longitudinal magnetic fields, $|\langle B_{\rm z} \rangle|_{\rm max}$ and $|\langle B_{\rm z} \rangle|_{\rm mean}$, were first obtained from a set of theoretical spectra computed for 20 rotational phases and then averaged over 100 different random realizations of the magnetic field maps with the same $r_{\rm sp}$ and $f_{\rm sp}$.

\begin{figure}[!t]
\centering
\figps{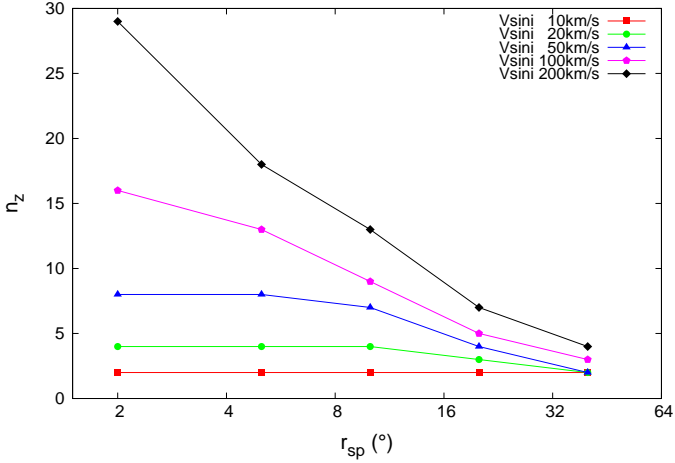}
\caption{Number of intersections of the zero line in the Stokes $V$ profiles as a function of the magnetic spot radius and projected rotational velocity. Each point is a median computed over 20 rotational phases and 100 different random realizations of magnetic spot distributions.}
\label{fig:res2}
\end{figure}

\section{Results}
\label{results}

Our main set of the line profile calculations for different $r_{\rm sp}$ and \vs\ was generated assuming \teff\,=\,20\,000~K, $i=60\degr$, and $f_{\rm sp}=0.5$. Here we present results obtained with this combination of parameters and then assess the impact of changing the effective temperature, inclination angle, and the filling factor. In all cases calculations were performed for $B_{\rm sp}=500$~G.

The mean longitudinal magnetic field and the amplitude of the Stokes $V$ profiles are plotted as a function of the magnetic spot size and projected rotational velocity in Fig.~\ref{fig:res1}. We illustrate the maximum \bzm\ and the largest polarization amplitude that might be expected for the observations with a dense phase coverage (left column) and the phase-averaged values of these observables, representative of snapshot observations (right column).

As one can see from the upper panels of Fig.~\ref{fig:res1}, the maximum longitudinal magnetic field changes from $\sim$\,10~G for $r_{\rm sp}=2\degr$ to 150--180~G for $r_{\rm sp}=40\degr$. The corresponding $|\langle B_{\rm z}\rangle|_{\rm mean}$ is approximately two times smaller. There is a small variation of \bzm\ with \vs, which is not expected for the purely geometrical definition of this magnetic parameter. This \vs\ dependence is real and is explained by the presence of weak structures in the local LSD Stokes spectra outside main line profiles (see Fig.~\ref{fig:prf_teff}). These structures contribute differently to the disk-integrated profiles depending on the range of rotational Doppler shifts, resulting in slightly different first-order moments of $\langle V_{\rm LSD}\rangle$ for different \vs. We have verified that the \vs\ dependence of \bzm\ disappears if we substitute numerical local LSD profile tables with (less realistic) smooth analytical profiles.

The lower panels in Fig.~\ref{fig:res1} illustrate the mean and maximum Stokes $V$ amplitude as a function of $r_{\rm sp}$ and \vs. This figure reveals a non-trivial interplay between the Doppler resolution and cancellation of the opposite spot polarities. For sharp-lined stars there is a strong dependence of the polarization amplitude on $r_{\rm sp}$, with the largest spots yielding about 10 times higher amplitude of the Zeeman polarization signatures. As soon as \vs\ exceeds $\sim$\,100~\kms, the amplitude of polarization profiles becomes almost independent of $r_{\rm sp}$. The typical peak-to-peak line polarization amplitude obtained in our numerical experiments lies in the range from $\sim$\,10$^{-3}$ to $\sim$\,10$^{-4}$.

According to our results, the low-\vs\ stars show a higher amplitude polarization features independently of the typical magnetic spot size. In other words, an increase of the polarization amplitude due to the rotational Doppler resolution does not compensate the reduction of the signal due to weakening of spectral lines. This suggests that massive sharp-lined stars (intrinsically slow rotators or fast rotators visible pole-on) are preferred targets for searching for signatures of complex magnetic fields. However, it should be remembered that an observational assessment of the typical sizes of magnetic features is only possible for fast rotators. To this end, Fig.~\ref{fig:res2} presents an analysis of the Stokes $V$ profile complexity using the number of zero line intersections, $n_{\rm z}$, for the same range of \vs\ and $r_{\rm sp}$ as shown in Fig.~\ref{fig:res1}. Clearly, a high \vs\ is beneficial for ascertaining the field complexity and resolving the surface structures at the smallest spatial scales. At the same time, a moderate \vs\ of $\sim$\,50~\kms\ is already sufficient for distinguishing, with a single high-quality spectropolarimetric observation, small-scale magnetic fields from the global fossil magnetic topologies.

All calculations presented here assumed 500~G field strength inside magnetic spots. Our results can be easily scaled to an arbitrary starspot field strength by multiplying the magnetic observables by $B_{\rm sp}/500$ (for the field strength $B_{\rm sp}$ measured in G). For the logarithmic plots shown in Fig.~\ref{fig:res1} this re-scaling corresponds to a vertical translation of all curves by $\lg (B_{\rm sp}/500)$. For example, decreasing the spot field strength to 100~G will result in the downward shift of the Stokes $V$ profile amplitude curves by a factor of $\approx$\,0.7. The morphology of the circular polarization profiles and Fig.~\ref{fig:res2} will remain unchanged. 

In the following we describe the impact of variation of \teff, $i$, and $f_{\rm sp}$ with respect to the default values of these parameters (\teff\,=\,20\,000~K, $i=60\degr$, and $f_{\rm sp}=0.5$). First, repeating the full set of calculations for $i=30\degr$ and $90\degr$, we found no dependence of any of the considered Stokes $V$ profile characteristics on the stellar inclination. Statistically, the observational signatures of localized magnetic fields turn out to be completely insensitive to this parameter.

The impact of using the local line profile tables corresponding to different \teff\ can be qualitatively assessed with the help of the local Stokes $I$ and $V$ profiles shown in Fig.~\ref{fig:prf_teff}. Compared to the full \teff\ range of 10\,000--30\,000~K, \teff\,=\,20\,000~K adopted for our main set of calculations represents the least favorable case due to the smallest Stokes $I$ line depth. Unsurprisingly, using \teff\,=\,10\,000~K increases the $\langle V_{\rm LSD} \rangle$ amplitude by 30\% for rapidly rotating stars and up to 70\% for sharp-lined stars. The change associated with increasing \teff\ from 20\,000 to 30\,000~K is less significant. In this case, an increase of the line depth is counteracted by an increase of the local profile width. As a result, we obtained $\approx$\,20\% Stokes $V$ signal enhancement for sharp-lined stars and no change or even a few per cent decrease of the polarization signal for rapid rotators.

Finally, we investigated the impact of reducing the filling factor of the magnetic spots from 0.5 to 0.2. As expected, this large decrease of the amount of unsigned magnetic flux on the stellar surface yields some reduction of both the Stokes $V$ amplitude and the mean longitudinal magnetic field inferred from the circular polarization profiles. However, this reduction is not linear with $f_{\rm sp}$ because a smaller number of magnetic spots on the visible hemisphere of the star also leads to a less efficient cancellation of the opposite magnetic field polarities. On average, we found about 50\% decrease of $V_{\rm max}$, $V_{\rm mean}$ and as well as $|\langle B_{\rm z} \rangle|_{\rm max}$ and $|\langle B_{\rm z} \rangle|_{\rm mean}$ for a factor of 2.5 reduction of the surface area covered with magnetic spots.

\section{Discussion}
\label{disc}

\begin{figure}[!t]
\centering
\figps{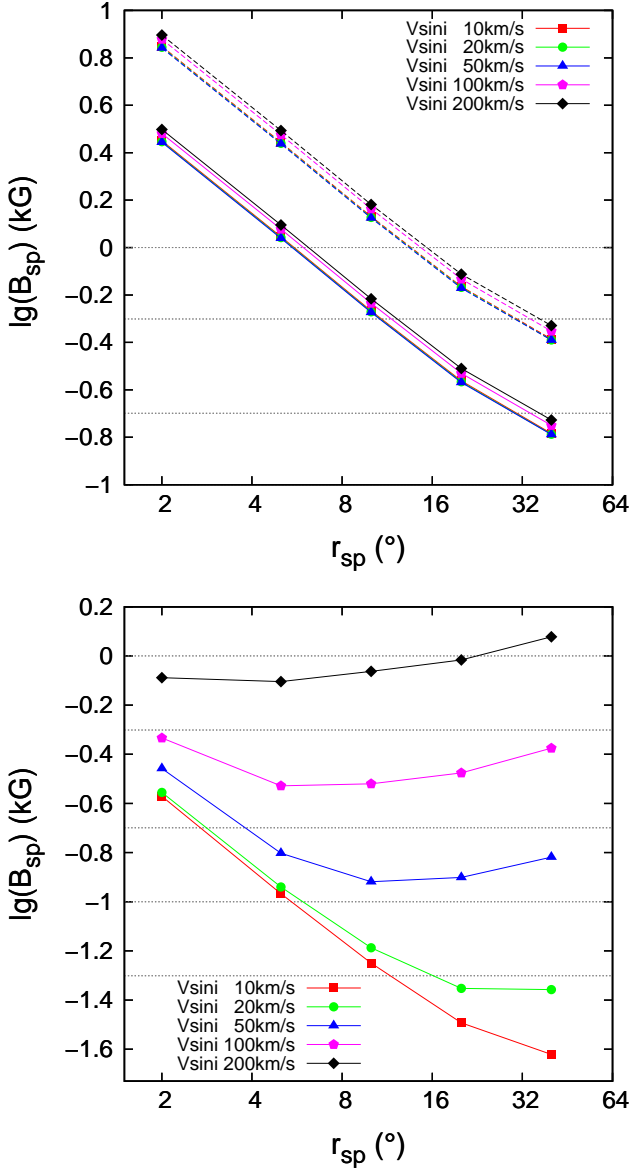}
\caption{Upper limits for the small-scale magnetic field intensity corresponding to the null results of the MiMeS survey. Upper panel: $B_{\rm sp}$ corresponding to \bzm\,=\,30~G (\textit{solid lines}) and 75~G (\textit{dashed lines}). Lower panel: $B_{\rm sp}$ required for the magnetic field detection with the LSD Stokes $V$ profiles at a confidence level of $\ge$\,99.9\% for at least half of the rotational phases. The dotted horizontal lines show $B_{\rm sp}=50$, 100, 200, 500, and 1000~G.}
\label{fig:res3}
\end{figure}

In the previous section we calculated the amplitudes of the mean longitudinal magnetic field and the circular polarization inside spectral lines for a fixed magnetic field strength and for different magnetic spot sizes. These predictions cannot be directly compared to any observations because no detection of small-scale magnetic fields has been reported so far for early-B and O stars. Nevertheless, it is of interest to interpret our results in terms of the maximum strength of such fields compatible with the most sensitive of the existing observational non-detections.

The most comprehensive search of the magnetic fields in massive stars was carried out by the MiMeS survey using the ESPaDOnS, NARVAL, and HARPSpol spectropolarimeters. Several discoveries of the strong, large-scale magnetic fields have been made from these observations \citep[e.g.][]{grunhut:2009,alecian:2011a,wade:2012a}. In addition, the MiMeS survey component (SC) include over 300 O and B stars for which no evidence of magnetic field was found. According to \citet{wade:2012b}, the resulting incidence of global magnetic fields is about 6\% for both O and B stars.

A summary of the magnetic field non-detections for the MiMeS SC stars was kindly communicated to us by J.~Grunhut. The median uncertainty of the \bz\ measurements is 25~G for all SC stars. Somewhat lower error bars of $\sim$\,10~G were achieved for about two dozen stars with \vs\,$\le$\,100~\kms. The median signal-to-noise ratio of the LSD Stokes $V$ profiles is 23\,000. Using these observational constraints, we inferred the upper limits for the magnetic field strength using our line profile and \bzm\ predictions for \teff\,=\,20\,000~K, $f_{\rm sp}=0.5$, and $i=60\degr$. 

The upper panel of Fig.~\ref{fig:res3} shows the minimum $B_{\rm sp}$ necessary for a $3\sigma$ detection of the mean longitudinal magnetic field for $\sigma=25$~G and 10~G. Evidently, \bz\ is not a particularly useful diagnostic when it comes to small-scale magnetic field topologies. The field detection using \bz\ requires $B_{\rm sp}\ge 500$~G for a subgroup of the MiMeS SC targets with the most accurate longitudinal magnetic field measurements. This limit is $B_{\rm sp}\ga 1500$~G for the entire SC sample.

Next, we assessed the ability of the MiMeS survey to detect small-scale magnetic fields with the LSD line profile analysis. We conducted a series of Monte-Carlo simulations by adding a random noise with different amplitude to the theoretical LSD Stokes $V$ profiles corresponding to $r_{\rm sp}=10\degr$ and estimating the probability of polarization signal detection using the chi-square statistics \citep{donati:1992}. These calculations showed that, independently of the Stokes $V$ profile morphology, the circular polarization signatures can be detected at a confidence level of $\ge99.9$\% for at least half of the rotational phases if the noise is at least five times smaller than the peak-to-peak amplitude of the Stokes $V$ profile. Combining this estimate with a typical precision of the Stokes $V$ LSD profiles obtained by MiMeS, we could infer the maximum possible strength of the magnetic field inside small-scale spots. This $B_{\rm sp}$ limit is illustrated in the lower panel of Fig.~\ref{fig:res3} as a function of $r_{\rm sp}$ and \vs. This figure confirms that the LSD Stokes $V$ profile analysis is considerably more sensitive to the small-scale fields compared to \bz. We find that for sharp-lined stars the detection of $B_{\rm sp}$\,$\approx$\,$50$~G field is possible for $r_{\rm sp}>10\degr$. The limiting field strength increases to 200--300~G for the smallest spatial scales we have considered ($r_{\rm sp}=2\degr$). Fast-rotating stars show a less pronounced dependence of the maximum $B_{\rm sp}$ on $r_{\rm sp}$. Typically, one should be able to detect the signatures of 300--500~G magnetic fields for any $r_{\rm sp}$ at \vs\,=\,100~\kms\ and 800--1200~G fields at \vs\,=\,200~\kms. Interestingly, for fast rotating stars the limit of 800~G corresponds to the field with $r_{\rm sp}\le5\degr$ and it is comparatively more difficult to detect larger-scale magnetic field structures.

How do these constraints compare with theoretical predictions? The only quantitative estimate of the strength of localized magnetic fields emerging on the surfaces of massive stars was presented by \citet[][hereafter CB11]{cantiello:2011} in the context of their study of the effects of turbulence in the iron SCZ. Supposing that the dynamo action in the convection zone is in equipartition with the kinetic energy of the gas, the authors found magnetic fields of up to $\sim$\,2~kG at the base of the surface radiative zone. Further assuming that the magnetic fields are brought up to the surface by the magnetic buoyancy, CB11 obtained the magnetic field strength of the order of 20--150~G in the photospheric layers, depending on the magnetic flux tube geometry and stellar parameters. CB11 also did not rule out the photospheric fields reaching the equipartition limit of about 300~G.

Little is known about the typical sizes of the surface magnetic field structures generated by the SCZ. CB11 argued that the lower limit of the magnetic starspot size is given by the pressure scale height, which is about 0.2\,$R_\odot$ for a massive star with $R_\star=10R_\odot$. This corresponds to $r_{\rm sp}\approx1\degr$, i.e. close to the lower spatial limit investigated in our paper. At the same time, magneto-hydrodynamic simulations demonstrate formation of magnetic structures on the scales significantly larger than the scale of convection \citep{kapyla:2008,cantiello:2011a}. For the Sun, the observed active regions also have sizes significantly exceeding the pressure scale height \citep[e.g.][]{zhang:2010}. Inhomogeneous wind models by \citep{cranmer:1996} considered surface spots with $r_{\rm sp}\approx10\degr$, hence the magnetic concentrations with coherent field orientation should be similarly large for the magnetic interpretation of CIRs and DACs. Taking into account these arguments, we suggest that $r_{\rm sp}=5$--10\degr\ might be a more realistic estimate of the typical size of the hypothetical surface magnetic field regions.

A remarkable result of our analysis is that the high-resolution Stokes $V$ observations of massive sharp-lined stars have already reached the precision necessary for detecting $r_{\rm sp}=5$--10\degr\ spots, filled with $B_{\rm sp}=50$--100~G radial magnetic field and occupying half of the stellar surface. However, no discovery of a corresponding population of narrow-line magnetic massive stars have been reported by MiMeS. This may indicate that the theoretical predictions by CB11 overestimate the surface field strength or that the fraction of the stellar surface covered by magnetic regions is much smaller than $f_{\rm sp}=0.5$ adopted in our simulations. A more firm conclusion emerging from our work is the absence of equipartition fields with $B_{\rm sp}\approx300$~G on the surfaces of massive stars. 

The tentative comparison of the overall MiMeS survey results with the generic theoretical predictions for a large range of stellar parameters can be put on more solid footing by considering fundamental parameters, observational constraints, and theoretical predictions for individual stars. On the one hand, such analysis would take into account the individual line lists, \vs, and circular polarization detection limits and, on the other hand, would allow choosing an appropriate maximum theoretical magnetic field strength according to the stellar luminosity and temperature (see Fig.~3 in CB11). In the future, it will be straightforward to perform this analysis of individual massive stars with the tools and methodology developed in our paper.

\begin{acknowledgements}
We thank Dr. Jason Grunhut for providing a summary of the MiMeS null results ahead of publication.
OK is a Royal Swedish Academy of Sciences Research Fellow, supported by the grants from Knut and Alice Wallenberg Foundation and Swedish Research Council. NS is supported by the Saint-Petersburg State University through a research grant 6.38.73.2011.
\end{acknowledgements}


\begin{thebibliography}{54}
\expandafter\ifx\csname natexlab\endcsname\relax\def\natexlab#1{#1}\fi

\bibitem[{{Alecian} {et~al.}(2011){Alecian}, {Kochukhov}, {Neiner}, {Wade}, {de
  Batz}, {Henrichs}, {Grunhut}, {Bouret}, {Briquet}, {Gagne}, {Naze}, {Oksala},
  {Rivinius}, {Townsend}, {Walborn}, {Weiss}, \& {Mimes
  Collaboration}}]{alecian:2011a}
{Alecian}, E., {Kochukhov}, O., {Neiner}, C., {et~al.} 2011, \aap, 536, L6

\bibitem[{{Auri{\`e}re} {et~al.}(2007){Auri{\`e}re}, {Wade}, {Silvester},
  {Ligni{\`e}res}, {Bagnulo}, {Bale}, {Dintrans}, {Donati}, {Folsom},
  {Gruberbauer}, {Bon Hoa}, {Jeffers}, {Johnson}, {Landstreet}, {L{\`e}bre},
  {Lueftinger}, {Marsden}, {Mouillet}, {Naseri}, {Paletou}, {Petit}, {Power},
  {Rincon}, {Strasser}, \& {Toqu{\'e}}}]{auriere:2007}
{Auri{\`e}re}, M., {Wade}, G.~A., {Silvester}, J., {et~al.} 2007, \aap, 475,
  1053

\bibitem[{{Bagnulo} {et~al.}(2012){Bagnulo}, {Landstreet}, {Fossati}, \&
  {Kochukhov}}]{bagnulo:2012}
{Bagnulo}, S., {Landstreet}, J.~D., {Fossati}, L., \& {Kochukhov}, O. 2012,
  \aap, 538, A129

\bibitem[{{Bagnulo} {et~al.}(2002){Bagnulo}, {Szeifert}, {Wade}, {Landstreet},
  \& {Mathys}}]{bagnulo:2002a}
{Bagnulo}, S., {Szeifert}, T., {Wade}, G.~A., {Landstreet}, J.~D., \& {Mathys},
  G. 2002, \aap, 389, 191

\bibitem[{{Braithwaite} \& {Cantiello}(2013)}]{braithwaite:2013}
{Braithwaite}, J. \& {Cantiello}, M. 2013, \mnras, 428, 2789

\bibitem[{{Braithwaite} \& {Nordlund}(2006)}]{braithwaite:2006}
{Braithwaite}, J. \& {Nordlund}, {\AA}. 2006, \aap, 450, 1077

\bibitem[{{Brun} {et~al.}(2005){Brun}, {Browning}, \& {Toomre}}]{brun:2005}
{Brun}, A.~S., {Browning}, M.~K., \& {Toomre}, J. 2005, \apj, 629, 461

\bibitem[{{Cantiello} \& {Braithwaite}(2011)}]{cantiello:2011}
{Cantiello}, M. \& {Braithwaite}, J. 2011, \aap, 534, A140

\bibitem[{{Cantiello} {et~al.}(2011){Cantiello}, {Braithwaite}, {Brandenburg},
  {Del Sordo}, {K{\"a}pyl{\"a}}, \& {Langer}}]{cantiello:2011a}
{Cantiello}, M., {Braithwaite}, J., {Brandenburg}, A., {et~al.} 2011, in IAU
  Symposium, Vol. 272, IAU Symposium, ed. C.~{Neiner}, G.~{Wade}, G.~{Meynet},
  \& G.~{Peters}, 32--37

\bibitem[{{Cantiello} {et~al.}(2009){Cantiello}, {Langer}, {Brott}, {de Koter},
  {Shore}, {Vink}, {Voegler}, {Lennon}, \& {Yoon}}]{cantiello:2009}
{Cantiello}, M., {Langer}, N., {Brott}, I., {et~al.} 2009, \aap, 499, 279

\bibitem[{{Cranmer} \& {Owocki}(1996)}]{cranmer:1996}
{Cranmer}, S.~R. \& {Owocki}, S.~P. 1996, \apj, 462, 469

\bibitem[{{Degroote} {et~al.}(2010){Degroote}, {Briquet}, {Auvergne},
  {Sim{\'o}n-D{\'{\i}}az}, {Aerts}, {Noels}, {Rainer}, {Hareter}, {Poretti},
  {Mahy}, {Oreiro}, {Vu{\v c}kovi{\'c}}, {Smolders}, {Baglin}, {Baudin},
  {Catala}, {Michel}, \& {Samadi}}]{degroote:2010}
{Degroote}, P., {Briquet}, M., {Auvergne}, M., {et~al.} 2010, \aap, 519, A38

\bibitem[{{Donati} {et~al.}(2002){Donati}, {Babel}, {Harries}, {Howarth},
  {Petit}, \& {Semel}}]{donati:2002}
{Donati}, J.-F., {Babel}, J., {Harries}, T.~J., {et~al.} 2002, \mnras, 333, 55

\bibitem[{{Donati} {et~al.}(2003){Donati}, {Cameron}, {Semel}, {Hussain},
  {Petit}, {Carter}, {Marsden}, {Mengel}, {L{\'o}pez Ariste}, {Jeffers}, \&
  {Rees}}]{donati:2003}
{Donati}, J.-F., {Cameron}, A.~C., {Semel}, M., {et~al.} 2003, \mnras, 345,
  1145

\bibitem[{{Donati} \& {Landstreet}(2009)}]{donati:2009}
{Donati}, J.-F. \& {Landstreet}, J.~D. 2009, \araa, 47, 333

\bibitem[{{Donati} {et~al.}(1997){Donati}, {Semel}, {Carter}, {Rees}, \&
  {Collier Cameron}}]{donati:1997}
{Donati}, J.-F., {Semel}, M., {Carter}, B.~D., {Rees}, D.~E., \& {Collier
  Cameron}, A. 1997, \mnras, 291, 658

\bibitem[{{Donati} {et~al.}(1992){Donati}, {Semel}, \& {Rees}}]{donati:1992}
{Donati}, J.-F., {Semel}, M., \& {Rees}, D.~E. 1992, \aap, 265, 669

\bibitem[{{Fullerton} {et~al.}(1996){Fullerton}, {Gies}, \&
  {Bolton}}]{fullerton:1996}
{Fullerton}, A.~W., {Gies}, D.~R., \& {Bolton}, C.~T. 1996, \apjs, 103, 475

\bibitem[{{Grunhut} {et~al.}(2009){Grunhut}, {Wade}, {Marcolino}, {Petit},
  {Henrichs}, {Cohen}, {Alecian}, {Bohlender}, {Bouret}, {Kochukhov}, {Neiner},
  {St-Louis}, \& {Townsend}}]{grunhut:2009}
{Grunhut}, J.~H., {Wade}, G.~A., {Marcolino}, W.~L.~F., {et~al.} 2009, \mnras,
  400, L94

\bibitem[{{Heger} {et~al.}(2005){Heger}, {Woosley}, \& {Spruit}}]{heger:2005}
{Heger}, A., {Woosley}, S.~E., \& {Spruit}, H.~C. 2005, \apj, 626, 350

\bibitem[{{Henrichs} {et~al.}(2005){Henrichs}, {Schnerr}, \& {ten
  Kulve}}]{henrichs:2005}
{Henrichs}, H.~F., {Schnerr}, R.~S., \& {ten Kulve}, E. 2005, in Astronomical
  Society of the Pacific Conference Series, Vol. 337, The Nature and Evolution
  of Disks Around Hot Stars, ed. R.~{Ignace} \& K.~G. {Gayley}, 114

\bibitem[{{Hubrig} {et~al.}(2011){Hubrig}, {Sch{\"o}ller}, {Kharchenko},
  {Langer}, {de Wit}, {Ilyin}, {Kholtygin}, {Piskunov}, {Przybilla}, \& {Magori
  Collaboration}}]{hubrig:2011}
{Hubrig}, S., {Sch{\"o}ller}, M., {Kharchenko}, N.~V., {et~al.} 2011, \aap,
  528, A151

\bibitem[{{Hunter} {et~al.}(2008){Hunter}, {Lennon}, {Dufton}, {Trundle},
  {Sim{\'o}n-D{\'{\i}}az}, {Smartt}, {Ryans}, \& {Evans}}]{hunter:2008}
{Hunter}, I., {Lennon}, D.~J., {Dufton}, P.~L., {et~al.} 2008, \aap, 479, 541

\bibitem[{{Kaper} {et~al.}(1996){Kaper}, {Henrichs}, {Nichols}, {Snoek},
  {Volten}, \& {Zwarthoed}}]{kaper:1996}
{Kaper}, L., {Henrichs}, H.~F., {Nichols}, J.~S., {et~al.} 1996, \aaps, 116,
  257

\bibitem[{{K{\"a}pyl{\"a}} {et~al.}(2008){K{\"a}pyl{\"a}}, {Korpi}, \&
  {Brandenburg}}]{kapyla:2008}
{K{\"a}pyl{\"a}}, P.~J., {Korpi}, M.~J., \& {Brandenburg}, A. 2008, \aap, 491,
  353

\bibitem[{{Kochukhov}(2011)}]{kochukhov:2011c}
{Kochukhov}, O. 2011, in IAU Symposium, Vol. 273, IAU Symposium, ed. D.~{Prasad
  Choudhary} \& K.~G. {Strassmeier}, 249--255

\bibitem[{{Kochukhov} {et~al.}(2010){Kochukhov}, {Makaganiuk}, \&
  {Piskunov}}]{kochukhov:2010a}
{Kochukhov}, O., {Makaganiuk}, V., \& {Piskunov}, N. 2010, \aap, 524, A5

\bibitem[{{Kochukhov} {et~al.}(2013){Kochukhov}, {Mantere}, {Hackman}, \&
  {Ilyin}}]{kochukhov:2013}
{Kochukhov}, O., {Mantere}, M.~J., {Hackman}, T., \& {Ilyin}, I. 2013, \aap,
  550, A84

\bibitem[{{Kupka} {et~al.}(1999){Kupka}, {Piskunov}, {Ryabchikova}, {Stempels},
  \& {Weiss}}]{kupka:1999}
{Kupka}, F., {Piskunov}, N., {Ryabchikova}, T.~A., {Stempels}, H.~C., \&
  {Weiss}, W.~W. 1999, \aaps, 138, 119

\bibitem[{{Kurucz}(1993)}]{kurucz:1993c}
{Kurucz}, R. 1993, ATLAS9 Stellar Atmosphere Programs and 2 km/s grid.~Kurucz
  CD-ROM No.~13.~ Cambridge, Mass.: Smithsonian Astrophysical Observatory.

\bibitem[{{Ligni{\`e}res} {et~al.}(2009){Ligni{\`e}res}, {Petit}, {B{\"o}hm},
  \& {Auri{\`e}re}}]{lignieres:2009}
{Ligni{\`e}res}, F., {Petit}, P., {B{\"o}hm}, T., \& {Auri{\`e}re}, M. 2009,
  \aap, 500, L41

\bibitem[{{MacGregor} \& {Cassinelli}(2003)}]{macgregor:2003}
{MacGregor}, K.~B. \& {Cassinelli}, J.~P. 2003, \apj, 586, 480

\bibitem[{{Maeder} \& {Meynet}(2005)}]{maeder:2005}
{Maeder}, A. \& {Meynet}, G. 2005, \aap, 440, 1041

\bibitem[{{Martins} {et~al.}(2010){Martins}, {Donati}, {Marcolino}, {Bouret},
  {Wade}, {Escolano}, \& {Howarth}}]{martins:2010}
{Martins}, F., {Donati}, J., {Marcolino}, W.~L.~F., {et~al.} 2010, \mnras, 407,
  1423

\bibitem[{{Mathys}(2009)}]{mathys:2009}
{Mathys}, G. 2009, in Astronomical Society of the Pacific Conference Series,
  Vol. 405, Solar Polarization 5: In Honor of Jan Stenflo, ed. S.~V.
  {Berdyugina}, K.~N. {Nagendra}, \& R.~{Ramelli}, 473

\bibitem[{{Mullan}(1984)}]{mullan:1984}
{Mullan}, D.~J. 1984, \apj, 283, 303

\bibitem[{{Mullan} \& {MacDonald}(2005)}]{mullan:2005}
{Mullan}, D.~J. \& {MacDonald}, J. 2005, \mnras, 356, 1139

\bibitem[{{Neiner} {et~al.}(2012){Neiner}, {Grunhut}, {Petit}, {ud-Doula},
  {Wade}, {Landstreet}, {de Batz}, {Cochard}, {Guti{\'e}rrez-Soto}, \&
  {Huat}}]{neiner:2012}
{Neiner}, C., {Grunhut}, J.~H., {Petit}, V., {et~al.} 2012, \mnras, 426, 2738

\bibitem[{{Petit} {et~al.}(2011){Petit}, {Ligni{\`e}res}, {Auri{\`e}re},
  {Wade}, {Alina}, {Ballot}, {B{\"o}hm}, {Jouve}, {Oza}, {Paletou}, \&
  {Th{\'e}ado}}]{petit:2011}
{Petit}, P., {Ligni{\`e}res}, F., {Auri{\`e}re}, M., {et~al.} 2011, \aap, 532,
  L13

\bibitem[{{Petit} \& {Wade}(2012)}]{petit:2012}
{Petit}, V. \& {Wade}, G.~A. 2012, \mnras, 420, 773

\bibitem[{{Piskunov} \& {Kochukhov}(2002)}]{piskunov:2002a}
{Piskunov}, N. \& {Kochukhov}, O. 2002, \aap, 381, 736

\bibitem[{{Prinja} {et~al.}(2002){Prinja}, {Massa}, \&
  {Fullerton}}]{prinja:2002}
{Prinja}, R.~K., {Massa}, D., \& {Fullerton}, A.~W. 2002, \aap, 388, 587

\bibitem[{{Semel}(1989)}]{semel:1989}
{Semel}, M. 1989, \aap, 225, 456

\bibitem[{{Shultz} {et~al.}(2012){Shultz}, {Wade}, {Grunhut}, {Bagnulo},
  {Landstreet}, {Neiner}, {Alecian}, {Hanes}, \& {MiMeS
  Collaboration}}]{shultz:2012}
{Shultz}, M., {Wade}, G.~A., {Grunhut}, J., {et~al.} 2012, \apj, 750, 2

\bibitem[{{Spruit}(2002)}]{spruit:2002}
{Spruit}, H.~C. 2002, \aap, 381, 923

\bibitem[{{Wade} {et~al.}(2011{\natexlab{a}}){Wade}, {Alecian}, {Bohlender},
  {Bouret}, {Cohen}, {Duez}, {Gagn{\'e}}, {Grunhut}, {Henrichs}, {Hill},
  {Kochukhov}, {Mathis}, {Neiner}, {Oksala}, {Owocki}, {Petit}, {Shultz},
  {Rivinius}, {Townsend}, {Vink}, \& {Vink}}]{wade:2011}
{Wade}, G.~A., {Alecian}, E., {Bohlender}, D.~A., {et~al.} 2011{\natexlab{a}},
  in IAU Symposium, Vol. 272, IAU Symposium, ed. C.~{Neiner}, G.~{Wade},
  G.~{Meynet}, \& G.~{Peters}, 118--123

\bibitem[{{Wade} {et~al.}(2007){Wade}, {Bagnulo}, {Drouin}, {Landstreet}, \&
  {Monin}}]{wade:2007}
{Wade}, G.~A., {Bagnulo}, S., {Drouin}, D., {Landstreet}, J.~D., \& {Monin}, D.
  2007, \mnras, 376, 1145

\bibitem[{{Wade} {et~al.}(2006){Wade}, {Fullerton}, {Donati}, {Landstreet},
  {Petit}, \& {Strasser}}]{wade:2006b}
{Wade}, G.~A., {Fullerton}, A.~W., {Donati}, J.-F., {et~al.} 2006, \aap, 451,
  195

\bibitem[{{Wade} {et~al.}(2012{\natexlab{a}}){Wade}, {Grunhut}, {Gr{\"a}fener},
  {Howarth}, {Martins}, {Petit}, {Vink}, {Bagnulo}, {Folsom}, {Naz{\'e}},
  {Walborn}, {Townsend}, \& {Evans}}]{wade:2012}
{Wade}, G.~A., {Grunhut}, J., {Gr{\"a}fener}, G., {et~al.} 2012{\natexlab{a}},
  \mnras, 419, 2459

\bibitem[{{Wade} {et~al.}(2012{\natexlab{b}}){Wade}, {Grunhut}, \& {MiMeS
  Collaboration}}]{wade:2012b}
{Wade}, G.~A., {Grunhut}, J.~H., \& {MiMeS Collaboration}. 2012{\natexlab{b}},
  in Astronomical Society of the Pacific Conference Series, Vol. 464,
  Circumstellar Dynamics at High Resolution, ed. A.~C. {Carciofi} \&
  T.~{Rivinius}, 405

\bibitem[{{Wade} {et~al.}(2011{\natexlab{b}}){Wade}, {Howarth}, {Townsend},
  {Grunhut}, {Shultz}, {Bouret}, {Fullerton}, {Marcolino}, {Martins},
  {Naz{\'e}}, {Ud Doula}, {Walborn}, \& {Donati}}]{wade:2011a}
{Wade}, G.~A., {Howarth}, I.~D., {Townsend}, R.~H.~D., {et~al.}
  2011{\natexlab{b}}, \mnras, 416, 3160

\bibitem[{{Wade} {et~al.}(2012{\natexlab{c}}){Wade}, {Ma{\'{\i}}z
  Apell{\'a}niz}, {Martins}, {Petit}, {Grunhut}, {Walborn}, {Barb{\'a}},
  {Gagn{\'e}}, {Garc{\'{\i}}a-Melendo}, {Jose}, {Moffat}, {Naz{\'e}}, {Neiner},
  {Pellerin}, {Penad{\'e}s Ordaz}, {Shultz}, {Sim{\'o}n-D{\'{\i}}az}, \&
  {Sota}}]{wade:2012a}
{Wade}, G.~A., {Ma{\'{\i}}z Apell{\'a}niz}, J., {Martins}, F., {et~al.}
  2012{\natexlab{c}}, \mnras, 425, 1278

\bibitem[{{Zahn} {et~al.}(2007){Zahn}, {Brun}, \& {Mathis}}]{zahn:2007}
{Zahn}, J.-P., {Brun}, A.~S., \& {Mathis}, S. 2007, \aap, 474, 145

\bibitem[{{Zhang} {et~al.}(2010){Zhang}, {Wang}, \& {Liu}}]{zhang:2010}
{Zhang}, J., {Wang}, Y., \& {Liu}, Y. 2010, \apj, 723, 1006

\end{thebibliography}

\end{document}